\documentclass[letter]{aa} % for the letters 
\usepackage{graphicx}
\usepackage{longtable}
\usepackage{lscape}
\usepackage{natbib}
\usepackage{lscape}
\usepackage[varg]{txfonts}
\begin{document}
   \title{GTC OSIRIS $z$-band imaging of Y dwarfs}

   \subtitle{ }

   \author{N. Lodieu \inst{1,2}\thanks{Based on
observations  made with the Gran Telescopio Canarias (GTC), installed in the
Spanish Observatorio del Roque de los Muchachos of the Instituto de
Astrof\'isica de Canarias, in the island of La Palma.}
          \and
          V.\ J.\ S.\ B\'ejar \inst{1,2}
          \and
          R.\ Rebolo \inst{1,2,3}
          }

   \institute{Instituto de Astrof\'isica de Canarias (IAC), Calle V\'ia L\'actea s/n, E-38200 La Laguna, Tenerife, Spain \\
         \email{nlodieu,vbejar,rrl@iac.es}
         \and
         Departamento de Astrof\'isica, Universidad de La Laguna (ULL),
E-38205 La Laguna, Tenerife, Spain
         \and
         Consejo Superior de Investigaciones Científicas, CSIC, Spain
             }

   \date{\today{}; \today{}}

% \abstract{}{}{}{}{} 
% 5 {} token are mandatory
 
  \abstract
   {}
   {The aim of the project is to contribute to the characterisation of 
the spectral energy distribution of the coolest brown dwarfs discovered 
to date, the Y dwarfs.}
   {We obtained $z$-band far-red imaging for six Y dwarfs and a
T9$+$Y0 binary with the OSIRIS (Optical System for Imaging and low 
Resolution Integrated Spectroscopy) instrument on the 10.4-m Gran 
Telescopio de Canarias (GTC).}
   {We detect five of the seven known Y dwarfs in the $z$-band, infer their
optical-to-infrared colours, and measure their proper motions. We find
a higher dispersion in the $z-J$ and $z-H$ colours of Y0 dwarfs 
than in T dwarfs. This dispersion is found to be correlated with $H-w2$.
%Although the uncertainties on the distances of Y dwarfs are large, we 
%find that the less luminous and probably the coolest Y0 dwarfs are the 
%bluest in both $J-H$ and $z-H$. 
The high dispersion in the optical-to-infrared colours of Y dwarfs
and the possible turn-over towards bluer colours may be a consequence
of the presence of sulfide clouds with different thicknesses,
the depletion of alcalines, and/or gravity effects.
% This points towards a transition from red to blue $z-H$ colours, 
% probably due to depletion of alcalines, which correlates well with the 
% increase in the $H-w2$ colour from late-T to Y dwarfs.
}
   {}


   \maketitle
%
%________________________________________________________________

%
%%%%%%%%%%%%%%%%%%%%%%%%%%%%%%
%%%%%  Introduction  %%%%%
%%%%%%%%%%%%%%%%%%%%%%%%%%%%%%
%
\section{Introduction}
\label{SDSSz_dY:intro}

The classical stellar sequence \citep{morgan43} has now been extended to 
include very cool objects that bridge the gap between brown dwarfs and planets, 
thanks to the advent of the Wide Infrared Survey Explorer 
\citep[WISE;][]{wright10} satellite. The ``Y'' class, originally proposed 
by \citet{kirkpatrick99}, now consists of 15 members 
\citep{cushing11,liu11a,kirkpatrick12,tinney12} with spectral types equal 
to, or later than, Y0 (tentative classification) and effective temperatures 
below 500\,K\@. Two additional candidates have been reported, but their
faintness currently hampers spectral typing \citep{liu12,luhman11}.

The number of late-T dwarfs with spectral types later than T6 has increased 
dramatically over the past few years thanks to large-scale surveys such as 
the UKIRT Infrared Deep Sky Survey 
\citep[UKIDSS;][]{lodieu07b,pinfield08,burningham09,burningham10a,burningham10b}, 
the Canada-France-Hawaii Brown Dwarf Survey 
\citep{delorme08a,delorme08b,reyle10,albert11}, and WISE
\citep{kirkpatrick12}. The T/Y transition is characterised by a rapid 
shift of the peak of the spectral energy distribution from the near-infrared to the mid-infrared 
and by narrower peaks in the $H$ band \citep{cushing11} with the possible
disappearance of the potassium absorption beyond 0.7 microns 
\citep{leggett12a}. Because of to their nature and low temperature, it is 
important to characterise their spectral energy distributions from the 
optical to the mid-infrared to understand the chemistry at 
play in these cool atmospheres.

Our goal is to contribute to the spectral energy characterisation of the Y 
dwarfs, pushing the limits of the Gran Telescopio de Canarias (GTC). This 
letter describes the optical imaging obtained for six Y dwarfs with OSIRIS 
\citep[Optical System for Imaging and low Resolution Integrated 
Spectroscopy;][]{cepa00} on the 10.4-m GTC\@. We present the imaging 
campaign and the associated data reduction and astrometry. We expose 
our results and place them into context with respect to the spectral 
energy distributions of T dwarfs \citep{leggett12a}.

\begin{table*}
 \centering
 \caption[]{GTC/OSIRIS astrometry, photometry (or 3$\sigma$ lower limits),
observing information, and photometry for six Y dwarfs and a T9$+$Y0
binary. The 
near-infrared photometry is taken from \citet{leggett12b}, except for the
$H$-band magnitude of WISEJ1828$+$2650 \citep{kirkpatrick12}.}
 \begin{tabular}{@{\hspace{0mm}}l @{\hspace{1mm}}c @{\hspace{1mm}}c @{\hspace{2mm}}c @{\hspace{2mm}}c @{\hspace{2mm}}c @{\hspace{2mm}}c @{\hspace{2mm}}c @{\hspace{2mm}}c @{\hspace{2mm}}c @{\hspace{2mm}}c@{\hspace{0mm}}}
 \hline
 \hline
WISE J\ldots{} (SpT) & R.A.         &     dec        &  $z$  &  Date & ExpT & $J$ & $H$ & $z-J$  & $\mu_\alpha\,cos\delta$ & $\mu_\delta$ \cr
 \hline
               & hh:mm:ss.sss & ${^\circ}$:':$''$ & mag & dd/mm/yy &  sec & mag &   mag \cr
 \hline
0146$+$4234    (Y0)  & 01:46:56.576 & $+$42:34:09.80 & 24.10$\pm$0.13 & 03/09/12 & 50$\times$50 & 19.40$\pm$0.25 & 18.71$\pm$0.24 & 5.39$\pm$0.27 & $-$0.52$\pm$0.08 & $-$0.11$\pm$0.08 \cr
% \hline
0410$+$1502    (Y0)  & 04:10:22.933 & $+$15:02:42.91 & 22.66$\pm$0.09 & 15/09/12 & 40$\times$50 & 19.44$\pm$0.03 & 20.02$\pm$0.05 & 3.22$\pm$0.09 & $+$1.20$\pm$0.08 & $-$2.17$\pm$0.08 \cr
% \hline
1405$+$5534    (Y0p)  &              &                & $>$23.85  & 06/08/12 & 45$\times$50 & 21.06$\pm$0.06 & 21.41$\pm$0.08 & $>$2.79 &  ---  &  --- \cr
% \hline
1738$+$2732 (Y0)$^{a}$  & 17:38:35.585 & $+$27:32:58.28 & 22.80$\pm$0.09 & 17/06/12 & 30$\times$50 & 20.05$\pm$0.09 & 20.45$\pm$0.09 & 2.75$\pm$0.13 & $+$0.32$\pm$0.10 & $-$0.39$\pm$0.10 \cr
% \hline
%1828$+$2650    ($\geq$Y2)  &              &                & $>$23.31--23.63  & 09/09/12 & 40$\times$50 & 23.57$\pm$0.35 & 22.85$\pm$0.24 & $>$$-$0.18--0.10 &  ---  &  --- \cr OB0005
%1828$+$2650    ($\geq$Y2)  &              &                & $>$24.33--24.59  & 06/08/12 & 39$\times$50 & 23.57$\pm$0.35 & 22.85$\pm$0.24 & $>$0.27--0.54 &  ---  &  --- \cr OB0005a
%1828$+$2650    ($\geq$Y2)  &              &                & $>$23.74--24.05  & 04/09/12 & 50$\times$50 & 23.57$\pm$0.35 & 22.85$\pm$0.24 & $>$0.17--0.48 &  ---  &  --- \cr OB0006
% 1828$+$2650    ($\geq$Y2)  &              &                & $>$24.02--24.30  & 09/09/12 & 40$\times$50 & 23.57$\pm$0.35 & 22.85$\pm$0.24 & $>$0.46--0.78 &  ---  &  --- \cr OB0007
1828$+$2650    ($\geq$Y2)  &              &                & $>$24.46  & 09/09/12 & 40$\times$50 & 23.48$\pm$0.23 & 22.85$\pm$0.24 & $>$0.98 &  ---  &  --- \cr
% \hline
2056$+$1459    (Y0)  & 20:56:29.028 & $+$14:59:54.64 & 23.09$\pm$0.08 & 15/06/12 & 40$\times$50 & 19.94$\pm$0.04 & 19.96$\pm$0.04 & 3.66$\pm$0.09 & $+$0.89$\pm$0.10 & $+$0.61$\pm$0.10 \cr
1217$+$1626 (T9+Y0)  & 12:17:57.144 & $+$16:26:35.99 & 21.60$\pm$0.03 & 15/12/12 & 15$\times$60 & 17.83$\pm$0.02 & 18.18$\pm$0.05 & 3.77$\pm$0.04 & $+$1.41$\pm$0.10 & $-$1.96$\pm$0.10 \cr
 \hline
 \label{tab_SDSSz_dY:phot_SDSSz}
 \end{tabular}
$^{a}$ We also targeted this object in the $i$ band with OSIRIS\@.
We measured a 3$\sigma$ lower limit of $i$\,$>$\,25.3 mag, 
translating into $i-z$\,$>$\,2.4 mag and $i-J$\,$>$\,5.8 mag.
\end{table*}
%

%
%%%%%%%%%%%%%%%%%%%%%%%%%%%%%%
%%%%% Optical imaging %%%%%
%%%%%%%%%%%%%%%%%%%%%%%%%%%%%%
%
\section{Far-red optical imaging}
\label{SDSSz_dY:obs}
\subsection{Observations}
\label{SDSSz_dY:obs_obs}

The OSIRIS instrument is mounted on the 10.4-m GTC operating 
at the Observatory del Roque de Los Muchachos (La Palma, Canary Islands).
The detector consists of two 2048$\times$4096 Marconi CCD42-82
separated by an 8 arcsec gap and operates at optical wavelengths,
from 365 to 1000 nm. The unvignetted instrument field-of-view is about
7 by 7 arcmin with an unbinned pixel scale of 0.125 arcsec.
We employed a 2$\times$2 binning because it is currently the 
standard mode of observations.

We imaged six of the seven Y dwarfs published by \citet{cushing11} 
and the T9+Y0 binary resolved by \citet{liu12} using 
the Sloan $z$ filter available on OSIRIS\@.
Table \ref{tab_SDSSz_dY:phot_SDSSz} provides the original names with the 
first four digits of the right ascension and declination as provided in 
\citet{cushing11} along with their current spectral types 
\citep{kirkpatrick12}.
This information is supplemented with the coordinates measured on the GTC 
OSIRIS images (when the target is detected), the $z$-band magnitudes with 
their associated error bars, the dates of observations, and the numbers of 
individual images with their respective integration times in seconds. 
Figure \ref{fig_SDSSz_dY:fc_dY} shows the GTC OSIRIS finding charts (two
arcmin a side) of all the observed objects (circled) with north up and east to the 
left. Table \ref{tab_SDSSz_dY:phot_SDSSz} also lists revised $J$ and 
$H$ magnitudes \citep{leggett12b} and the associated $z-J$ colours plotted 
in Fig.\ \ref{fig_SDSSz_dY:SpT_zmj_zmh_dY}.

The observations were conducted on 15 and 17 June, 6 and 13 August, 
3, 5, 9, and 16 September, and 15 December 2012\@. 
All observations were conducted under average seeing of 0.8--1.3 arcsec, 
photometric or clear conditions, grey time, and airmass less than 1.5\@. 
Bias and skyflats were obtained during the afternoon or morning of the 
respective nights. For each Y dwarf, we obtained three to five series of 
ten frames with 50 sec on-source individual integrations, except for the 
coolest \citep[WISE J1828$+$2650 tentatively classified as 
$>$Y2;][]{kirkpatrick12} for which we repeated the series four times to try 
to detect the object in the $z$-band and for WISE J1217$+$1626\,AB, 
for which we obtained three series of five frames of 60 sec.

\subsection{Data reduction and photometry}
\label{SDSSz_dY:obs_phot}

We reduced the OSIRIS Sloan $z$-band images in a standard manner with
IRAF\footnote{IRAF is distributed by the National Optical Astronomy
Observatories, which are operated by the Association of Universities for
Research in Astronomy, Inc., under cooperative agreement with the National
Science Foundation} \citep{tody93}. First, we subtracted the mean
bias and divided by the normalised averaged master skyflat to each 
individual science frame. Then, we combined each set of ten images taken 
without dithering, and finally we combined those sets, applying the offsets to 
create a master science frame. In the case of WISE J1217$+$1626\,AB, 
images were sky-subtracted before the combination.

We photometrically calibrated the $z$-band images with Sloan
standard fields \citep{smith02} or with Sloan images of the same 
field, using point sources with
photometric errors below 0.02--0.05 mag. We measured instrumental 
magnitudes for 5 to 10 stars in those fields and derived zero points
ranging from 27.976$\pm$0.037 to 28.091$\pm$0.049 mag.

\begin{figure*}
  \centering
  \includegraphics[width=0.19\linewidth, angle=0]{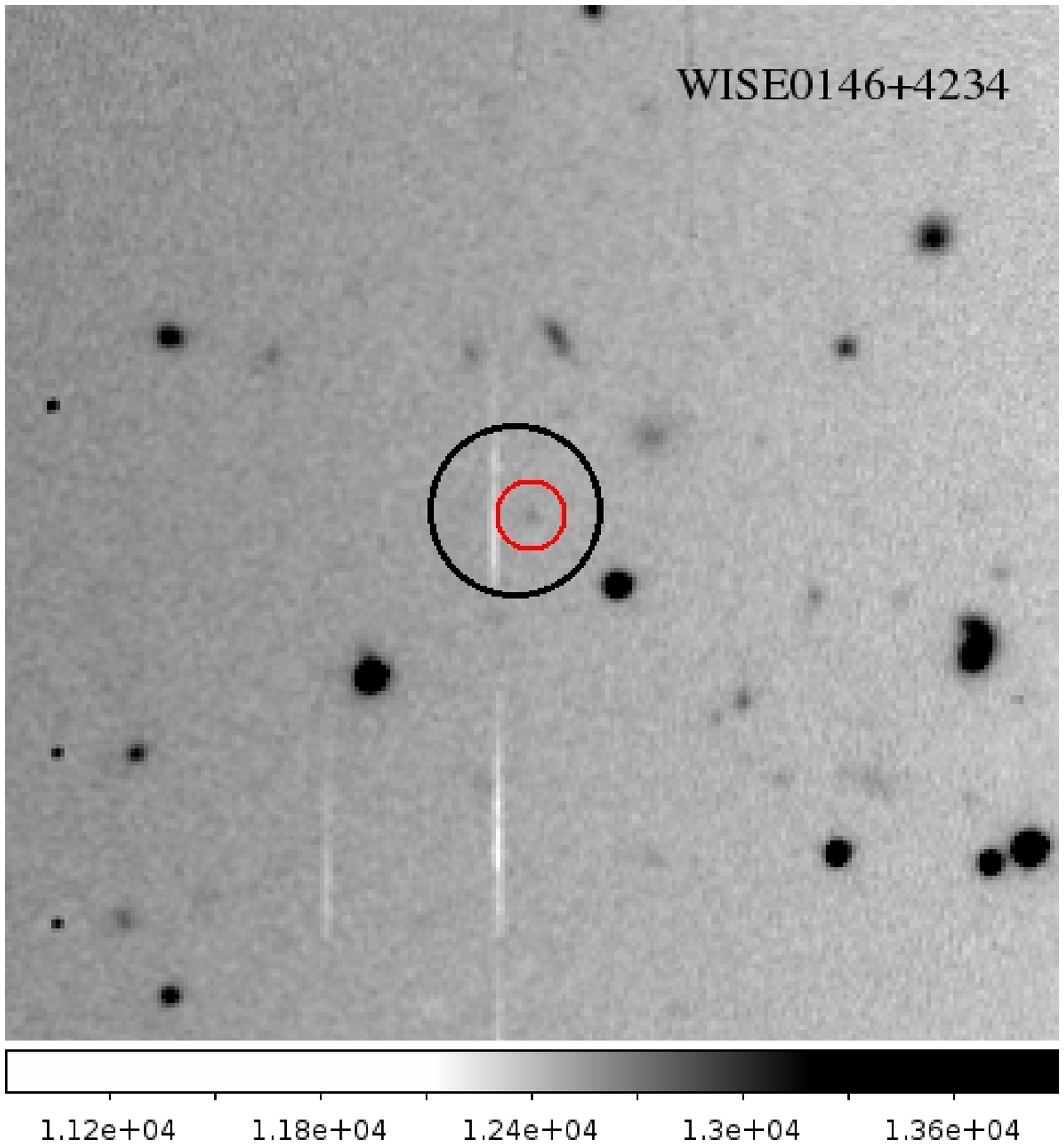}
  \includegraphics[width=0.19\linewidth, angle=0]{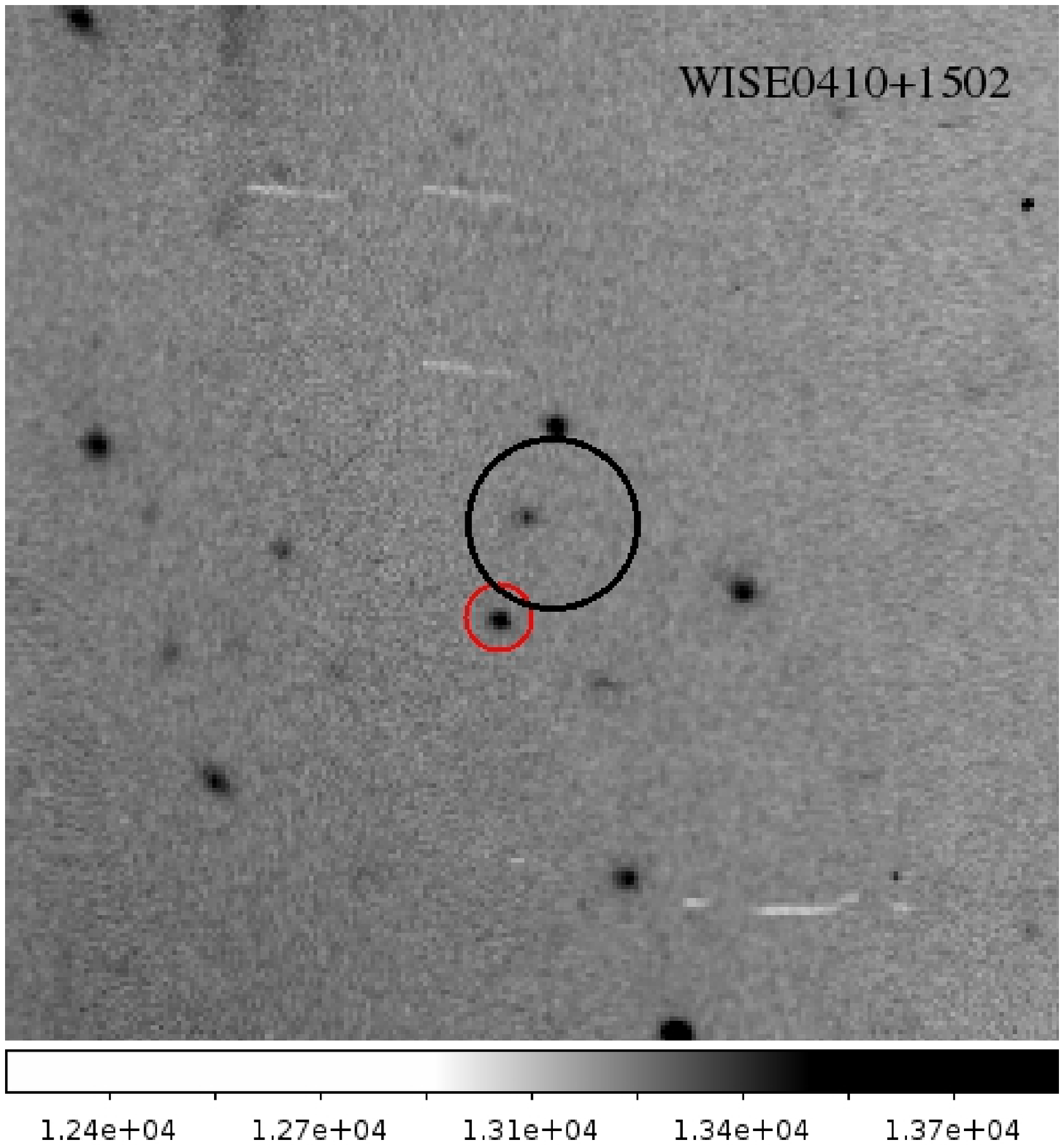}
  \includegraphics[width=0.19\linewidth, angle=0]{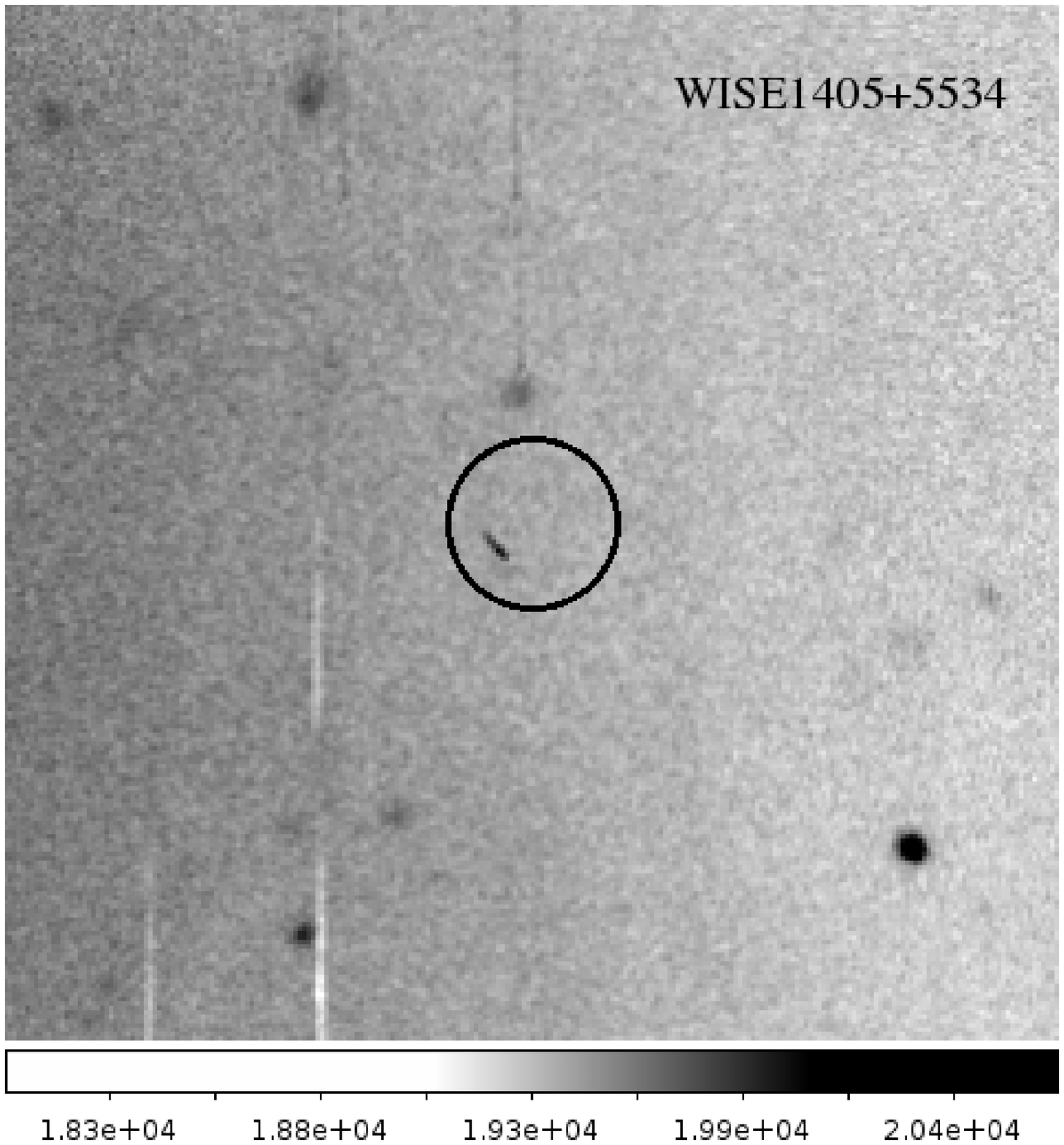}
%  \newline
  \includegraphics[width=0.19\linewidth, angle=0]{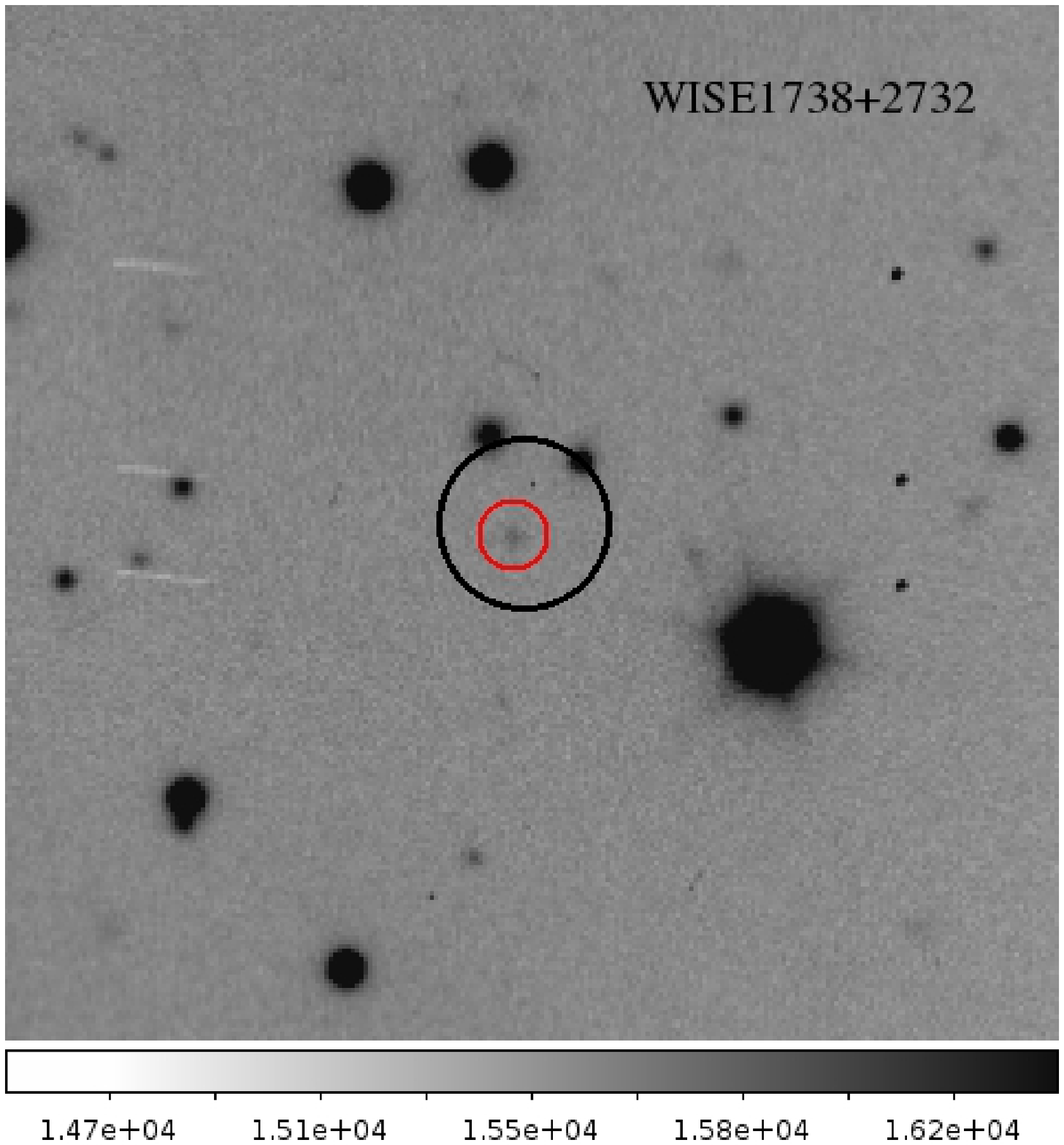}
  \includegraphics[width=0.19\linewidth, angle=0]{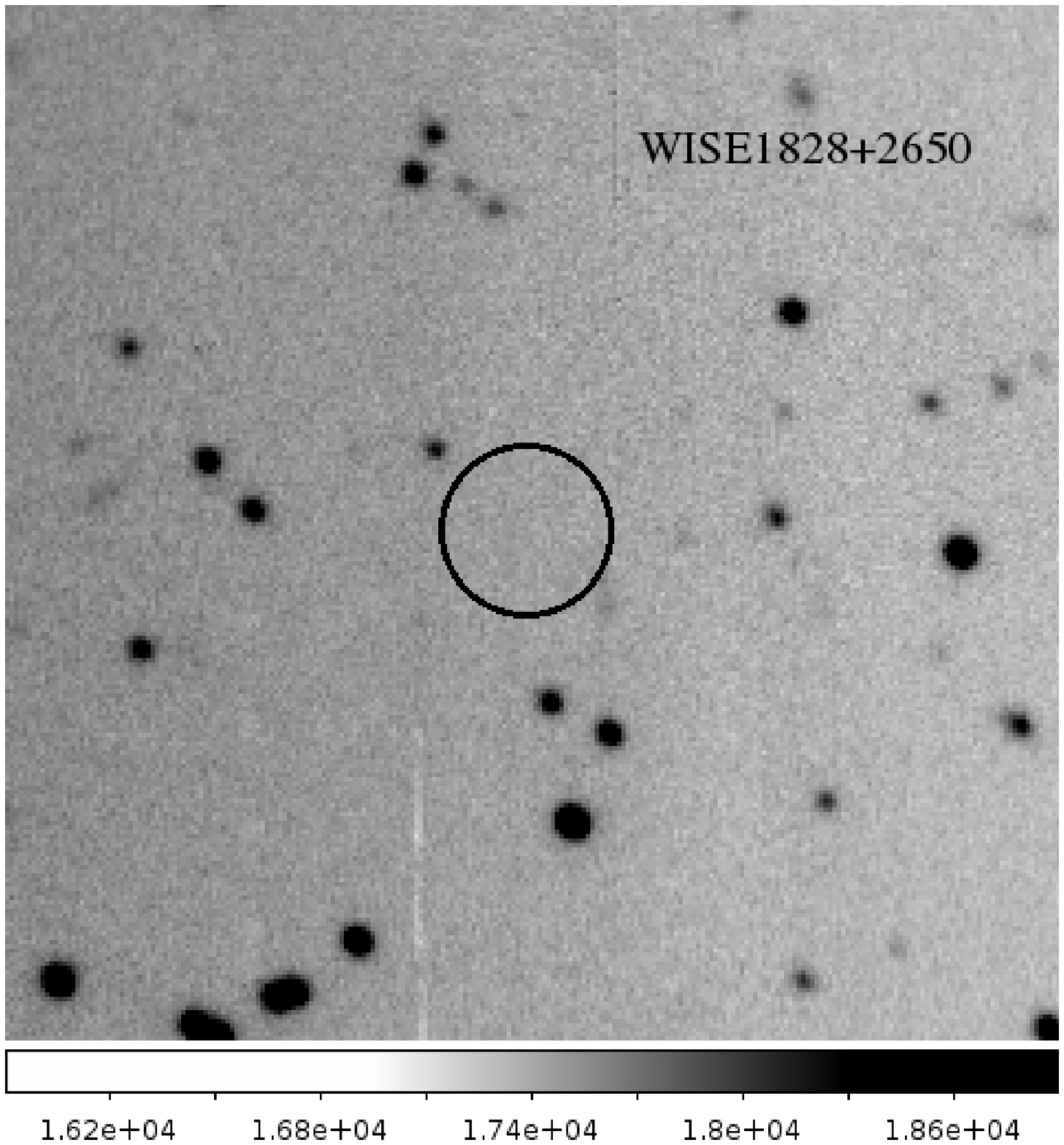}
  \includegraphics[width=0.19\linewidth, angle=0]{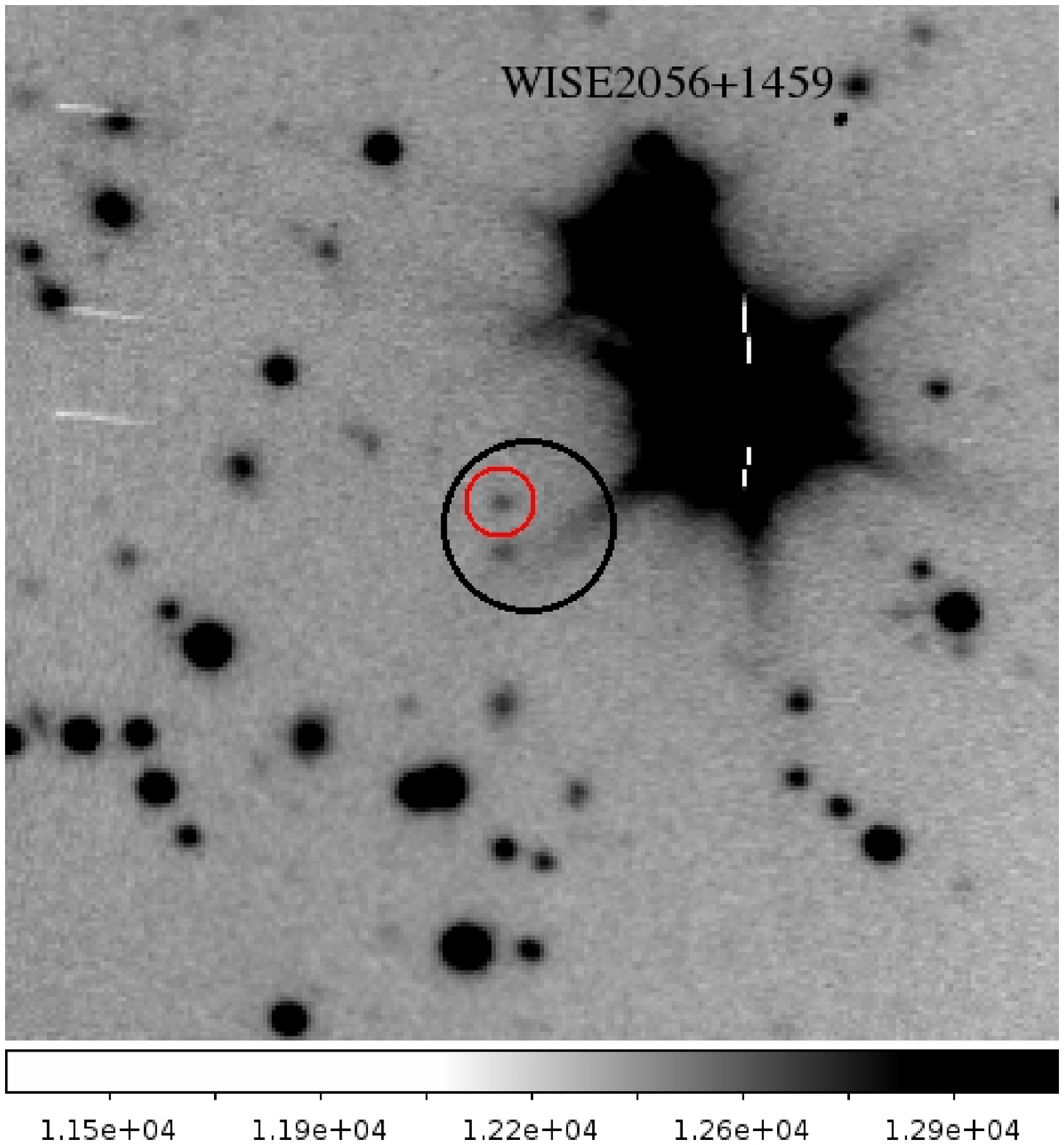}
  \includegraphics[width=0.19\linewidth, angle=0]{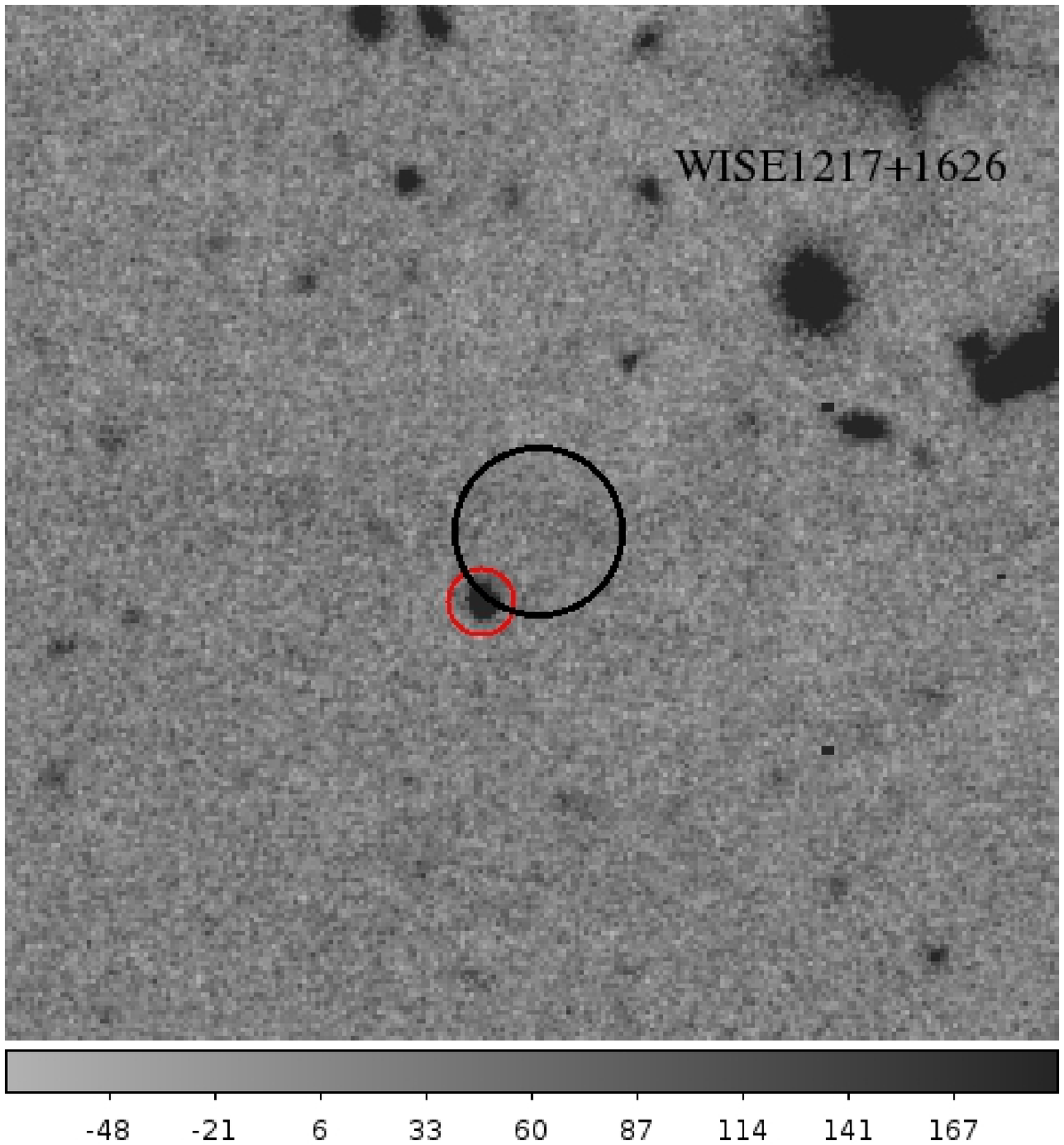}
   \caption{Finding charts for the six Y dwarfs and the T9$+$Y0 binary 
imaged in Sloan $z$ with GTC/OSIRIS\@. North is up and east is left. 
Images are 1$\times$1 arcmin Targets are ordered by increasing right 
ascension from left to right starting from the top left corner. 
The open circles have a radius of five arcsec and mark the WISE position 
of the Y dwarfs, showing the motion of the object 
when detected (small red circle with two arcsec radius).
   }
   \label{fig_SDSSz_dY:fc_dY}
\end{figure*}

We detected four of the six Y dwarfs and the T9+Y0 binary
(unresolved in our images) with OSIRIS in the $z$-band
(Table \ref{tab_SDSSz_dY:phot_SDSSz}; Fig.\ \ref{fig_SDSSz_dY:fc_dY}). 
We measured the photometry of the closest source to the nominal position
of the WISE source (Table \ref{tab_SDSSz_dY:phot_SDSSz}), except
for WISE J1410$+$1502 and WISE J1217$+$1626\,AB, which have moved 
beyond the five-arcsec circle drawn in Fig.\ \ref{fig_SDSSz_dY:fc_dY} 
due to their high proper motion ($\sim$2.5 arcsec/yr). To measure
the magnitudes, we performed aperture and point-spread function (PSF) 
photometry using {\tt{DAOPHOT}} in IRAF because of the presence of objects 
close to some of our targets. We chose an aperture equal to $\sim$3 times 
the full-width-at-half-maximum and checked that the target was 
well-subtracted by our PSF analysis without residuals. We transformed the 
instrumental magnitudes into apparent magnitudes using the zero-points 
derived for the photometric standard fields observed on the same night. 
The final calibrated magnitudes of the four Y dwarfs detected on the 
OSIRIS images are given in Table \ref{tab_SDSSz_dY:phot_SDSSz}.
For the remaining sources, we list the 3$\sigma$ lower limits
(Table \ref{tab_SDSSz_dY:phot_SDSSz}) computed from the root-mean-square of 
the sky at the nominal position of the target compared to the peak of the 
flux of three nearby (non-saturated and isolated) stars. After averaging
all images for WISE J1828$+$2650, we can set a lower limit of 
$z$\,$>$\,24.46 mag, translating into $z-H$\,$>$1.6 mag (and 
$z-J$\,$>$\,0.9 mag) for this (tentative) Y2 dwarf.

%
%%%%%%%%%%%%%%%%%%%%%%%%%%%%%%%
%%%%% Figure: Finding charts %%%%%
%%%%%%%%%%%%%%%%%%%%%%%%%%%%%%%
%
% Files saved as JPG from ds9 and then saved as PS with xv
% fc_WISE0146p4234_OSIRIS_z.ps
% fc_WISE0410p1502_OSIRIS_z.ps
% fc_WISE1405p5534_OSIRIS_z.ps
% fc_WISE1738p2732_OSIRIS_z.ps
% fc_WISE1828p2650_OSIRIS_z.ps
% fc_WISE2056p1459_OSIRIS_z.ps

%
\subsection{Astrometry}
\label{SDSSz_dY:obs_astrom}

We astrometrically calibrated the combined science frames with IRAF 
({\tt{daofind}}, {\tt{ccxymatch}}, and {\tt{ccmap}} routines)
using bright point sources from 2MASS 
\citep{cutri03} or SDSS DR8 in the case of WISE J1405p5534\@.
We typically found 50--100 stars (depending on crowding) in the OSIRIS
field-of-view, resulting in an astrometric calibration better than
0.1--0.15 arcsec. The GTC OSIRIS coordinates (right ascension and 
declination) of the five objects detected in the $z$-band are listed 
in columns 2 and 3 of Table \ref{tab_SDSSz_dY:phot_SDSSz}.
We measured proper motions, consistent with the values quoted by
\citet{marsh12} and \citet{liu12}, thanks to the 2.0--2.5 year baseline 
between the GTC and WISE observations. The error bars of each component 
of the proper motion take into account the WISE astrometric accuracy 
(0.15--0.17 arcsec) and conservative errors on the GTC centroid of 
0.15 arcsec (Table \ref{tab_SDSSz_dY:phot_SDSSz}).
We find high motions as expected for objects lying at distances
of only a few parsecs.

%
%%%%%%%%%%%%%%%%%%%%%%%%%%%%%%%
%%%%% Figure: z-J vs SpT %%%%%
%%%%%%%%%%%%%%%%%%%%%%%%%%%%%%%
%
% Plot (z-J,SpT) created with IDL program Tdwarf_colours_OSIRIS.pro
%
\begin{figure*}
  \centering
  \includegraphics[width=0.49\linewidth, angle=0]{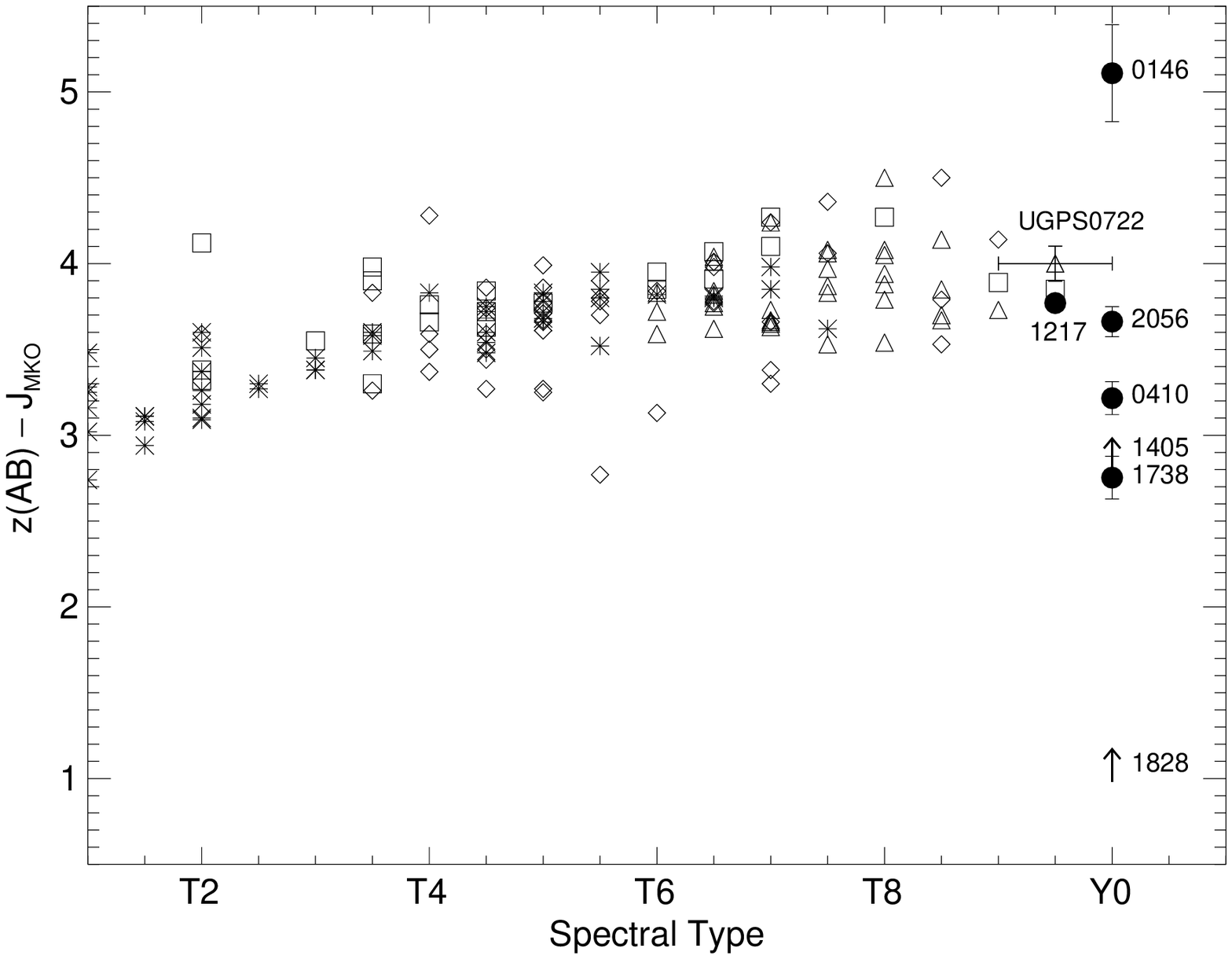}
  \includegraphics[width=0.49\linewidth, angle=0]{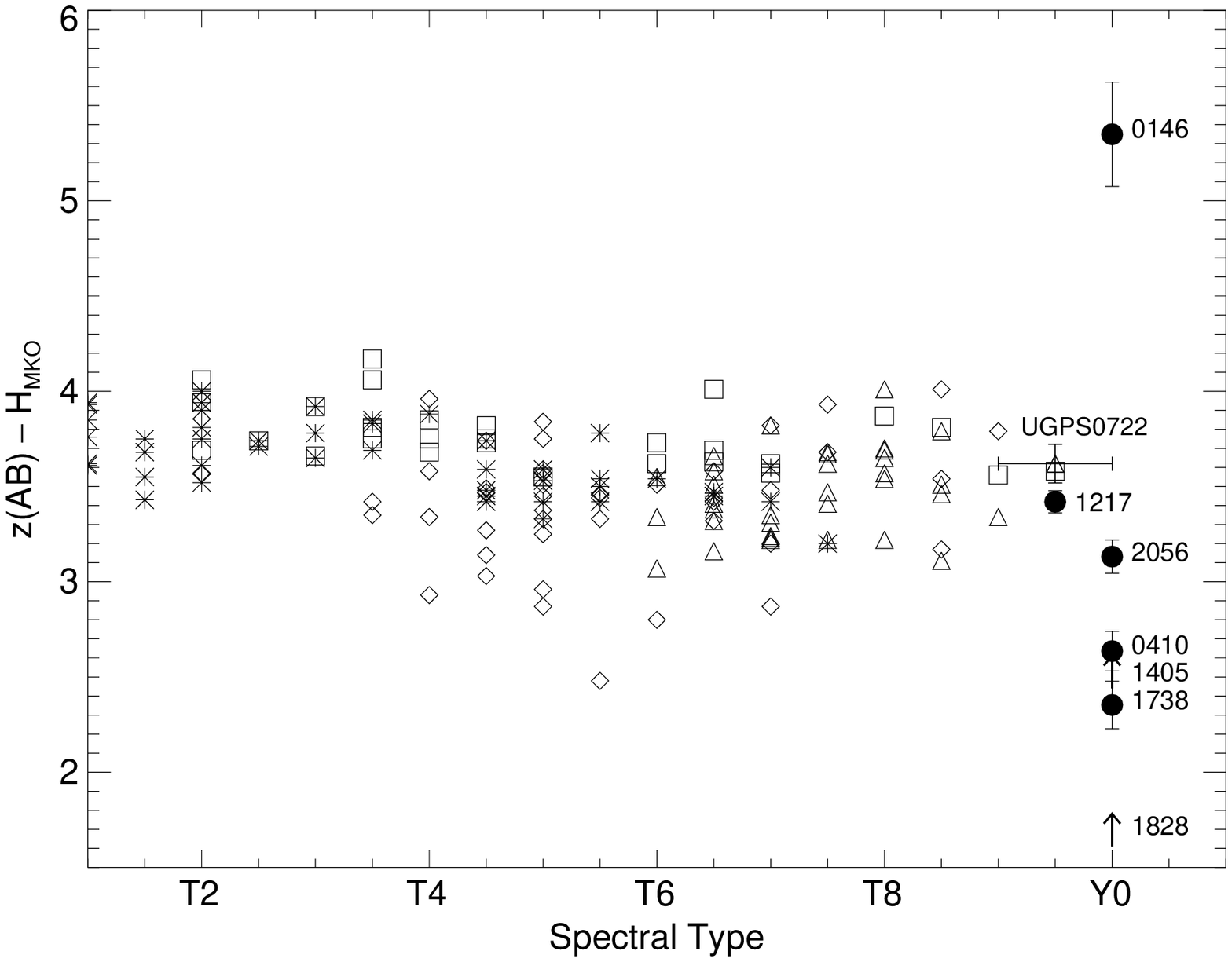}
   \caption{$z-J$ (left) and $z-H$ (right) colours of Y0
dwarfs (filled dots for detections and arrows for lower limits) as a 
function of spectral type compared to the colours of late-T dwarfs 
from UKIDSS \citep[diamonds;][]{burningham10b},
Sloan \citep[asterisks;][]{golimowski04a,knapp04,chiu06}, 
Canada-France brown dwarf survey \citep[squares;][]{albert11}, and 
the sample of \citet[][triangles]{leggett12b}.}
   \label{fig_SDSSz_dY:SpT_zmj_zmh_dY}
\end{figure*}
%

%
%%%%%%%%%%%%%%%%%%%%%%%%%%%%%%%
%%%%% Discussion  %%%%%
%%%%%%%%%%%%%%%%%%%%%%%%%%%%%%%
%
% Absolute magnitudes from distances in Kirkpatrick'12:
% 0146p4234 01:46:56.66 +42:34:10.0 H=18.71 errH=0.24 w2=15.08 d=6.3 (3.2-9.4) Mh=19.713 (18.844-21.184) Mw2=16.083 (15.214-17.554)
% 0410p1502 04:10:22.71 +15:02:48.4 H=19.05 errH=0.09 w2=14.18 d=6.1 (3.8-6.2) Mh=20.123 (20.088-21.151) Mw2=15.253 (15.218-16.281)
% 1738p2732 17:38:35.53 +27:32:59.0 H=20.39 errH=0.39 w2=14.55 d=9.0 (7.0-7.4) Mh=20.619 (21.044-21.164) Mw2=14.779 (15.204-15.324)
% 2056p1459 20:56:28.91 +14:59:53.2 H=19.62 errH=0.31 w2=13.93 d=5.2 (4.9-5.5) Mh=21.040 (20.918-21.169) Mw2=15.350 (15.228-15.479)
% UGPS0722  07:22:27.51 -05:04:31.2 H=16.90 errH=0.02 w2=12.17 d=4.12 (4.08-4.16) Mh=18.826 (18.805-18.848) Mw2=14.096 (14.075-14.117)
%
% Absolute magnitudes from parallaxes in Marsh'12:
% 0410p1502 04:10:22.71 +15:02:48.4 H=19.05 errH=0.09 w2=14.18 d=4.2 (3.6-5.4) Mh=17.166 (16.832-17.630) Mw2=12.296 (11.962-12.760)
% 1738p2732 17:38:35.53 +27:32:59.0 H=20.39 errH=0.39 w2=14.55 d>6.0 (>6.0   ) Mh>19.281 (XX.XXX-XX.XXX) Mw2>13.441 (XX.XXX-XX.XXX)
% 2056p1459 20:56:28.91 +14:59:53.2 H=19.62 errH=0.31 w2=13.93 d=7.5 (5.7-11.8) Mh=18.995 (18.399-19.979) Mw2=13.305 (12.709-14.289)
%
\section{Discussion and conclusions}
\label{SDSSz_dY:discussion}

We reported in this letter the first far-red optical detection of four Y0 
dwarfs and the unresolved T9+Y0 binary from \citet{liu12}, which are 
among the coolest brown dwarfs known to date. All four Y0 dwarfs with 
$z$-band detection have magnitudes in the range 22.8--24.1 mag, implying 
$z-J$ and $z-H$ colours spanning 2.7--5.2 mag and 
2.3--5.4 mag, respectively (Fig.\ \ref{fig_SDSSz_dY:SpT_zmj_zmh_dY}). 
We also presented lower limit of $z-J$ of 2.79 mag and 0.98 mag for
WISE J1405$+$5534 (T0pec?) and WISE J1828$+$2650 ($>$T2), respectively.

Figure \ref{fig_SDSSz_dY:SpT_zmj_zmh_dY} shows the $z-J$ and $z-H$ 
colours of the Y0 dwarfs as a function of spectral type (black dots with 
error bars and arrows for lower limits) along with the colours of T dwarfs.
Here we adopted a spectral type of T9.5+/-0.5 for the binary T9+Y0.
The dispersion observed for Y0 
dwarfs is greater than for T dwarfs. We added to this 
plot UGPS J072227.51$-$054031.2 classified as T10 by \citet{lucas10} 
and \citet{leggett12a}, but proposed as the T9 standard by \citet{cushing11}.

\begin{figure}
  \includegraphics[width=\linewidth, angle=0]{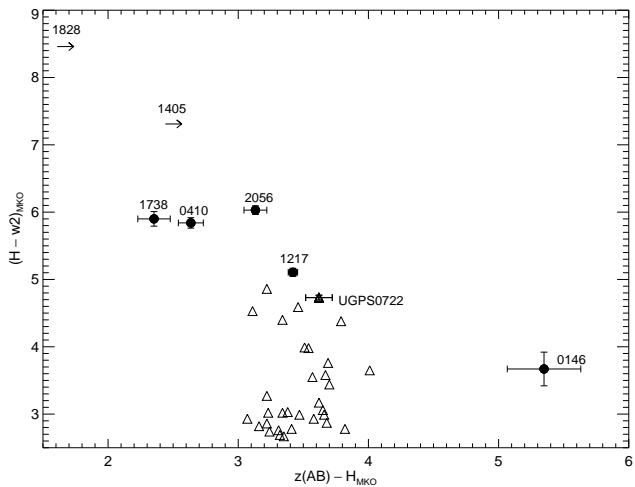}
   \caption{($z-H$,$H-w2$) colour-colour diagram for our targets (filled 
dots and arrows for lower limits), late-T dwarfs 
\citep[open triangles;][]{leggett12b}, and UGPS J0722227.51$-$054031.2 
\citep[open triangle with error bars;][]{lucas10}.}
   \label{fig_SDSSz_dY:zmh_hmw2_dY}
\end{figure}

Figure \ref{fig_SDSSz_dY:zmh_hmw2_dY} displays the ($z-H$,$H-w2$)
colour-colour diagram for the four Y0 dwarfs and the T9+Y0 binary
(filled dots) detected in 
$z$ and for the two sources with upper limits (left-pointing arrows).
This plot takes into account the improved near-infrared photometry
from \citet{leggett12b}, compared to the original values in Table 2
of \citet{kirkpatrick12}. We note a trend towards bluer colours with 
increasing $H-w2$ from late-T dwarfs to Y0 dwarfs, except for 
WISE J0146$+$4234, 
which looks peculiar in all diagrams. This trend seems to start off
around a spectral type of $\sim$T8 and agrees with the bluer 
$Y-J$ colours observed with cooler temperatures for the Y class and 
the excess of flux in the 840--940 nm region in the spectrum of 
WISE J2056$+$1459 compared to UGPS J072227.51$-$054031.2 
\citep{leggett12b}. The T9+Y0 binary from \citet{liu12} shows 
an intermediate position between late-T and Y0 dwarfs.
A possible explanation for the relative changes in the $z$-band may be 
associated to the disappearance of the strong atomic potassium and sodium 
bands into molecules (e.g.\ KCl$_{2}$ or NaCl) and possibly other alkalis 
as we move towards cooler temperatures.

%
%%%%%%%%%%%%%%%%%%%%%%%%%%%%%%%
%%%%% Figure: (z-H,H-w2) diagram %%%%%
%%%%%%%%%%%%%%%%%%%%%%%%%%%%%%%
%
% (z-J,H-w2) diagram created with the IDL program Tdwarf_colours_OSIRIS.pro
%
%

%
%%%%%%%%%%%%%%%%%%%%%%%%%%%%%%%
%%%%% Figure: (z-H,Y-H) diagram %%%%%
%%%%%%%%%%%%%%%%%%%%%%%%%%%%%%%
%
% (z-H,Y-H) diagram created with the IDL program Tdwarf_colours_OSIRIS.pro
%
\begin{figure}
  \includegraphics[width=\linewidth, angle=0]{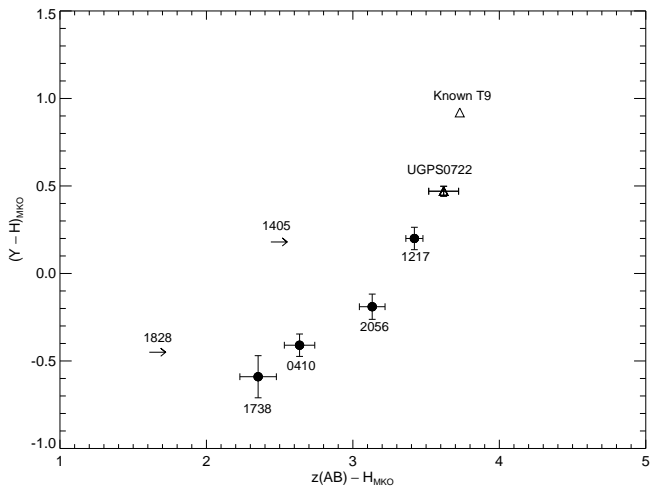}
   \caption{($z-H$,$Y-H$) colour-colour diagram for our targets (filled 
dots and arrows for lower limits), UGPS J0722227.51$-$054031.2 
\citep[open triangle with the error bars;][]{lucas10}, and the 
known T9, CFBDSIR\,145829.0$+$101343.0 \citep[open triangle;][]{delorme10}.}
   \label{fig_SDSSz_dY:zmh_ymh_dY}
\end{figure}

The spread in the $z-J$ and $z-H$ colours of Y0 dwarfs is
$\sim$2.5-3.0 mag, much wider than the spread observed throughout the
T dwarf sequence (Fig.\ \ref{fig_SDSSz_dY:SpT_zmj_zmh_dY}) and greater 
than the spread in the $J-K$ colours of L dwarfs \citep{hawley02}.
This spread can be qualitatively explained using recent models by
\citet{morley12}, which propose that the amount of sulfide clouds 
present in the atmospheres of these
Y dwarfs strongly impacts their spectral energy distributions (note that
this is not the case for iron or silicate clouds which play a role at
higher temperatures where the L/T transition takes place). Indeed,
Figure 5 of \citet{morley12} shows that the presence of sulfide clouds
decreases the emerging flux in the 0.8--1.3 microns with lower effective 
temperatures (starting off below $\sim$900\,K). Hence, the clouds
affect primarily $Y$ and $J$ but also the $z$-band, which is what we 
observe. Moreover, this effect seems to increase with decreasing 
sedimentation efficiency (the $f_{\rm sed}$ parameter in the models 
of \citet{marley02}). Similarly, Figure 10 of \citet{morley12} 
demonstrates that the combination of a wide range of gravities and 
presence of sulfide clouds with a variety sedimentation efficiencies 
increases the spread in the $J-H$ colours of brown dwarfs with 
temperatures below 600 K\@. These sulfide clouds may be responsible 
for the spread observed in the optical-to-infrared colours of Y dwarfs, 
suggesting that the atmosphere of WISE J0146 may contain thicker clouds 
and higher gravity than the other Y0 dwarfs in our sample. Table 3
of \citet{leggett12b} confirms that brown dwarfs with a temperature
of 400\,K influenced by thicker clouds would be fainter in $Y$ than 
those affected by thin clouds. As pointed out above, the same effect
could occur in the $z$ band.
Moreover, we observe a correlation between the $z-H$ and $Y-H$
colours of the three Y0 dwarfs (Fig.\ \ref{fig_SDSSz_dY:zmh_ymh_dY})
common to our sample and \citet{leggett12b}. The bluer objects in
$z-H$ turn out to be also rather blue in $Y-H$. If this trend holds, 
we would predict $Y-H$\,=\,0.96 mag for WISE J0146$+$4234\@.

Finally, it is important to complete the sample of $z$-band 
measurements for Y dwarfs in the southern hemisphere and obtain far-red 
optical spectroscopy, which is only available for WISE J2056$+$1459
\citep{leggett12b}.
This may show drastic changes as indicated by the $z$-band photometry,
which can probe changes in the atmospheric properties responsible for the
far-red spectral energy distribution of these brown dwarfs.
Our photometry would also help the modellers to reproduce
this part of the spectral energy distribution, disentangling the effects
of clouds, gravity, and pressure-broadened potassium and sodium bands.
Thirdly, the relatively blue optical-to-infrared colours of Y0
dwarfs and the possible turnover towards even bluer colours with lower
temperatures could have a significant impact on the strategies currently
used to identify Y dwarfs in large-scale surveys, such as UKIDSS,
CFBDS, Pan-Starrs \citep{deacon11}, and large synoptic survey telescope
\citep{lsst2009}.

%
%%%%%%%%%%%%%%%%%%%%%%%%%%%%%%%
%%%%%  ACKNOWLEDGEMENTS %%%%%
%%%%%%%%%%%%%%%%%%%%%%%%%%%%%%%
%
\begin{acknowledgements}
NL was funded by the Ram\'on y Cajal fellowship number 08-303-01-02\@.
This research has been supported by the Spanish Ministry of Economics and 
Competitiveness under the projects AYA2010-19136, AYA2010-21308-C3-02, 
AYA2010-21308-C03-03 and AYA2010-20535\@. We are very grateful to the GTC 
operation team and to Antonio Cabrera-Lavers for 
his support at all stages of this project.
We thank Jos\'e Alberto Rubi{\~n}o, Ricardo Genova Santos, Rafael Barrena,
and Angela Hempel for obtaining photometric calibration at the Isaac Newton
telescope, and Bartek Gauza, Mar\'{\i}a Rosa Zapatero 
Osorio and Karla Pe\~{n}a Ram\'{\i}rez 
for their help in obtaining some of the OSIRIS/GTC images.

This work is based on observations made with the Gran Telescopio Canarias
(GTC), operated on the island of La Palma in the Spanish Observatorio del
Roque de los Muchachos of the Instituto de Astrof\'isica de Canarias.
This research has been supported by the Spanish Ministry of Economy and
Competitiveness (MINECO) under the grant AYA2010-19136\@.

This research has made use of the Simbad and Vizier databases, operated
at the Centre de Donn\'ees Astronomiques de Strasbourg (CDS), and
of NASA's Astrophysics Data System Bibliographic Services (ADS).
\end{acknowledgements}
%

%
%%%%%%%%%%%%%%%%%%%%%%%%%%%%%%%%%%%%%%%%%%
%%%%%%%%  Bibliography  %%%%%%%%
%%%%%%%%%%%%%%%%%%%%%%%%%%%%%%%%%%%%%%%%%%
%
% \begin{thebibliography}{}
\bibliographystyle{aa}
\bibliography{../../AA/mnemonic,../../AA/biblio_old}

\begin{thebibliography}{35}
\expandafter\ifx\csname natexlab\endcsname\relax\def\natexlab#1{#1}\fi

\bibitem[{{Albert} {et~al.}(2011){Albert}, {Artigau}, {Delorme}, {Reyl{\'e}},
  {Forveille}, {Delfosse}, \& {Willott}}]{albert11}
{Albert}, L., {Artigau}, {\'E}., {Delorme}, P., {et~al.} 2011, AJ, 141, 203

\bibitem[{{Burningham} {et~al.}(2010{\natexlab{a}}){Burningham}, {Leggett},
  {Lucas}, {Pinfield}, {Smart}, {Day-Jones}, \& {8 co-authors}}]{burningham10a}
{Burningham}, B., {Leggett}, S.~K., {Lucas}, P.~W., {et~al.}
  2010{\natexlab{a}}, MNRAS, 404, 1952

\bibitem[{{Burningham} {et~al.}(2009){Burningham}, {Pinfield}, {Leggett},
  {Tinney}, {Liu}, {Homeier}, {West}, \& {13 co-authors}}]{burningham09}
{Burningham}, B., {Pinfield}, D.~J., {Leggett}, S.~K., {et~al.} 2009, MNRAS,
  395, 1237

\bibitem[{{Burningham} {et~al.}(2010{\natexlab{b}}){Burningham}, {Pinfield},
  {Lucas}, {Leggett}, {Deacon}, {Tamura}, {Tinney}, {Lodieu}, \& {11
  co-authors}}]{burningham10b}
{Burningham}, B., {Pinfield}, D.~J., {Lucas}, P.~W., {et~al.}
  2010{\natexlab{b}}, MNRAS, 406, 1885

\bibitem[{{Cepa} {et~al.}(2000){Cepa}, {Aguiar}, {Escalera},
  {Gonzalez-Serrano}, {Joven-Alvarez}, {Peraza}, {Rasilla}, {Rodriguez-Ramos},
  {Gonzalez}, {Cobos Duenas}, {Sanchez}, {Tejada}, {Bland-Hawthorn},
  {Militello}, \& {Rosa}}]{cepa00}
{Cepa}, J., {Aguiar}, M., {Escalera}, V.~G., {et~al.} 2000, in Society of
  Photo-Optical Instrumentation Engineers (SPIE) Conference Series, Vol. 4008,
  Society of Photo-Optical Instrumentation Engineers (SPIE) Conference Series,
  ed. {M.~Iye \& A.~F.~Moorwood}, 623--631

\bibitem[{{Chiu} {et~al.}(2006){Chiu}, {Fan}, {Leggett}, {Golimowski}, {Zheng},
  {Geballe}, {Schneider}, \& {Brinkmann}}]{chiu06}
{Chiu}, K., {Fan}, X., {Leggett}, S.~K., {et~al.} 2006, AJ, 131, 2722

\bibitem[{{Cushing} {et~al.}(2011){Cushing}, {Kirkpatrick}, {Gelino},
  {Griffith}, {Skrutskie}, {Mainzer}, {Marsh}, {Beichman}, {Burgasser},
  {Prato}, {Simcoe}, {Marley}, {Saumon}, {Freedman}, {Eisenhardt}, \&
  {Wright}}]{cushing11}
{Cushing}, M.~C., {Kirkpatrick}, J.~D., {Gelino}, C.~R., {et~al.} 2011, ApJ,
  743, 50

\bibitem[{{Cutri} {et~al.}(2003){Cutri}, {Skrutskie}, {van Dyk}, {Beichman},
  {Carpenter}, {Chester}, {Cambresy}, {Evans}, {Fowler}, {Gizis}, \& {15
  coauthors}}]{cutri03}
{Cutri}, R.~M., {Skrutskie}, M.~F., {van Dyk}, S., {et~al.} 2003, 2MASS All Sky
  Catalog of point sources, 2246

\bibitem[{{Deacon} {et~al.}(2011){Deacon}, {Liu}, {Magnier}, {Bowler},
  {Goldman}, {Redstone}, {Burgett}, {Chambers}, {Flewelling}, {Kaiser},
  {Lupton}, {Morgan}, {Price}, {Sweeney}, {Tonry}, {Wainscoat}, \&
  {Waters}}]{deacon11}
{Deacon}, N.~R., {Liu}, M.~C., {Magnier}, E.~A., {et~al.} 2011, AJ, 142, 77

\bibitem[{{Delorme} {et~al.}(2010){Delorme}, {Albert}, {Forveille}, {Artigau},
  {Delfosse}, {Reyl{\'e}}, {Willott}, {Bertin}, {Wilkins}, {Allard}, \&
  {Arzoumanian}}]{delorme10}
{Delorme}, P., {Albert}, L., {Forveille}, T., {et~al.} 2010, A\&A, 518, A39

\bibitem[{{Delorme} {et~al.}(2008{\natexlab{a}}){Delorme}, {Delfosse},
  {Albert}, {Artigau}, {Forveille}, {Reyl{\'e}}, {Allard}, {Homeier}, {Robin},
  {Willott}, {Liu}, \& {Dupuy}}]{delorme08a}
{Delorme}, P., {Delfosse}, X., {Albert}, L., {et~al.} 2008{\natexlab{a}}, A\&A,
  482, 961

\bibitem[{{Delorme} {et~al.}(2008{\natexlab{b}}){Delorme}, {Willott},
  {Forveille}, {Delfosse}, {Reyl{\'e}}, {Bertin}, {Albert}, {Artigau}, {Robin},
  {Allard}, {Doyon}, \& {Hill}}]{delorme08b}
{Delorme}, P., {Willott}, C.~J., {Forveille}, T., {et~al.} 2008{\natexlab{b}},
  A\&A, 484, 469

\bibitem[{{Golimowski} {et~al.}(2004){Golimowski}, {Leggett}, {Marley}, {Fan},
  {Geballe}, {Knapp}, {Vrba}, {Henden}, \& {11 authors}}]{golimowski04a}
{Golimowski}, D.~A., {Leggett}, S.~K., {Marley}, M.~S., {et~al.} 2004, AJ, 127,
  3516

\bibitem[{{Hawley} {et~al.}(2002){Hawley}, {Covey}, {Knapp}, {Golimowski},
  {Fan}, {Anderson}, {Gunn}, {Harris}, {Ivezi{\' c}}, {Long}, \& {22
  coauthors}}]{hawley02}
{Hawley}, S.~L., {Covey}, K.~R., {Knapp}, G.~R., {et~al.} 2002, AJ, 123, 3409

\bibitem[{{Kirkpatrick} {et~al.}(2012){Kirkpatrick}, {Gelino}, {Cushing},
  {Mace}, {Griffith}, {Skrutskie}, {Marsh}, {Wright}, {Eisenhardt}, {McLean},
  {Mainzer}, {Burgasser}, {Tinney}, {Parker}, \& {Salter}}]{kirkpatrick12}
{Kirkpatrick}, J.~D., {Gelino}, C.~R., {Cushing}, M.~C., {et~al.} 2012, ApJ,
  753, 156

\bibitem[{{Kirkpatrick} {et~al.}(1999){Kirkpatrick}, {Reid}, {Liebert},
  {Cutri}, {Nelson}, {Beichman}, {Dahn}, {Monet}, {Gizis}, \&
  {Skrutskie}}]{kirkpatrick99}
{Kirkpatrick}, J.~D., {Reid}, I.~N., {Liebert}, J., {et~al.} 1999, ApJ, 519,
  802

\bibitem[{{Knapp} {et~al.}(2004){Knapp}, {Leggett}, {Fan}, {Marley}, {Geballe},
  {Golimowski}, {Finkbeiner}, {Gunn}, , \& {21 co-authors}}]{knapp04}
{Knapp}, G.~R., {Leggett}, S.~K., {Fan}, X., {et~al.} 2004, AJ, 127, 3553

\bibitem[{{Leggett} {et~al.}(2012{\natexlab{a}}){Leggett}, {Morley}, {Marley},
  {Saumon}, {Fortney}, \& {Visscher}}]{leggett12b}
{Leggett}, S.~K., {Morley}, C.~V., {Marley}, M.~S., {et~al.}
  2012{\natexlab{a}}, ArXiv e-prints

\bibitem[{{Leggett} {et~al.}(2012{\natexlab{b}}){Leggett}, {Saumon}, {Marley},
  {Lodders}, {Canty}, {Lucas}, {Smart}, {Tinney}, {Homeier}, {Allard},
  {Burningham}, {Day-Jones}, {Fegley}, {Ishii}, {Jones}, {Marocco}, {Pinfield},
  \& {Tamura}}]{leggett12a}
{Leggett}, S.~K., {Saumon}, D., {Marley}, M.~S., {et~al.} 2012{\natexlab{b}},
  ApJ, 748, 74

\bibitem[{{Liu} {et~al.}(2011){Liu}, {Delorme}, {Dupuy}, {Bowler}, {Albert},
  {Artigau}, {Reyl{\'e}}, {Forveille}, \& {Delfosse}}]{liu11a}
{Liu}, M.~C., {Delorme}, P., {Dupuy}, T.~J., {et~al.} 2011, ApJ, 740, 108

\bibitem[{{Liu} {et~al.}(2012){Liu}, {Dupuy}, {Bowler}, {Leggett}, \&
  {Best}}]{liu12}
{Liu}, M.~C., {Dupuy}, T.~J., {Bowler}, B.~P., {Leggett}, S.~K., \& {Best},
  W.~M.~J. 2012, ApJ, 758, 57

\bibitem[{{Lodieu} {et~al.}(2007){Lodieu}, {Pinfield}, {Leggett}, {Jameson},
  {Mortlock}, {Warren}, \& {co-authors}}]{lodieu07b}
{Lodieu}, N., {Pinfield}, D.~J., {Leggett}, S.~K., {et~al.} 2007, MNRAS, 379,
  1423

\bibitem[{{LSST Science Collaboration} {et~al.}(2009){LSST Science
  Collaboration}, {Abell}, {Allison}, {Anderson}, {Andrew}, {Angel}, {Armus},
  {Arnett}, {Asztalos}, {Axelrod}, \& et~al.}]{lsst2009}
{LSST Science Collaboration}, {Abell}, P.~A., {Allison}, J., {et~al.} 2009,
  ArXiv e-prints

\bibitem[{{Lucas} {et~al.}(2010){Lucas}, {Tinney}, {Burningham}, {Leggett},
  {Pinfield}, {Smart}, {Jones}, {Marocco}, {Barber}, {Yurchenko}, {Tennyson},
  {Ishii}, {Tamura}, {Day-Jones}, {Adamson}, {Allard}, \& {Homeier}}]{lucas10}
{Lucas}, P.~W., {Tinney}, C.~G., {Burningham}, B., {et~al.} 2010, MNRAS, L124

\bibitem[{{Luhman} {et~al.}(2011){Luhman}, {Burgasser}, \&
  {Bochanski}}]{luhman11}
{Luhman}, K.~L., {Burgasser}, A.~J., \& {Bochanski}, J.~J. 2011, ApJL, 730, L9

\bibitem[{{Marley} {et~al.}(2002){Marley}, {Seager}, {Saumon}, {Lodders},
  {Ackerman}, {Freedman}, \& {Fan}}]{marley02}
{Marley}, M.~S., {Seager}, S., {Saumon}, D., {et~al.} 2002, ApJ, 568, 335

\bibitem[{{Marsh} {et~al.}(2012){Marsh}, {Wright}, {Kirkpatrick}, {Gelino},
  {Cushing}, {Griffith}, {Skrutskie}, \& {Eisenhardt}}]{marsh12}
{Marsh}, K.~A., {Wright}, E.~L., {Kirkpatrick}, J.~D., {et~al.} 2012, ArXiv
  e-prints

\bibitem[{{Morgan} {et~al.}(1943){Morgan}, {Keenan}, \& {Kellman}}]{morgan43}
{Morgan}, W.~W., {Keenan}, P.~C., \& {Kellman}, E. 1943, {An atlas of stellar
  spectra, with an outline of spectral classification} (Chicago, Ill., The
  University of Chicago press)

\bibitem[{{Morley} {et~al.}(2012){Morley}, {Fortney}, {Marley}, {Visscher},
  {Saumon}, \& {Leggett}}]{morley12}
{Morley}, C.~V., {Fortney}, J.~J., {Marley}, M.~S., {et~al.} 2012, ApJ, 756,
  172

\bibitem[{{Pinfield} {et~al.}(2008){Pinfield}, {Burningham}, {Tamura},
  {Leggett}, {Lodieu}, {Lucas}, {Mortlock}, \& {28 co-authors}}]{pinfield08}
{Pinfield}, D.~J., {Burningham}, B., {Tamura}, M., {et~al.} 2008, MNRAS, 390,
  304

\bibitem[{{Reyl{\'e}} {et~al.}(2010){Reyl{\'e}}, {Delorme}, {Willott},
  {Albert}, {Delfosse}, {Forveille}, {Artigau}, {Malo}, {Hill}, \&
  {Doyon}}]{reyle10}
{Reyl{\'e}}, C., {Delorme}, P., {Willott}, C.~J., {et~al.} 2010, A\&A, 522,
  A112

\bibitem[{{Smith} {et~al.}(2002){Smith}, {Tucker}, {Kent}, {Richmond},
  {Fukugita}, {Ichikawa}, {Ichikawa}, {Jorgensen}, {Uomoto}, {Gunn}, {Hamabe},
  {Watanabe}, {Tolea}, {Henden}, {Annis}, {Pier}, {McKay}, {Brinkmann}, {Chen},
  {Holtzman}, {Shimasaku}, \& {York}}]{smith02}
{Smith}, J.~A., {Tucker}, D.~L., {Kent}, S., {et~al.} 2002, AJ, 123, 2121

\bibitem[{{Tinney} {et~al.}(2012){Tinney}, {Faherty}, {Kirkpatrick}, {Wright},
  {Gelino}, {Cushing}, {Griffith}, \& {Salter}}]{tinney12}
{Tinney}, C.~G., {Faherty}, J.~K., {Kirkpatrick}, J.~D., {et~al.} 2012, ApJ,
  759, 60

\bibitem[{{Tody}(1993)}]{tody93}
{Tody}, D. 1993, in Astronomical Society of the Pacific Conference Series,
  Vol.~52, Astronomical Data Analysis Software and Systems II, ed. R.~J.
  {Hanisch}, R.~J.~V. {Brissenden}, \& J.~{Barnes}, 173

\bibitem[{{Wright} {et~al.}(2010){Wright}, {Eisenhardt}, {Mainzer}, {Ressler},
  {Cutri}, {Jarrett}, {Kirkpatrick}, \& {31 co-authors}}]{wright10}
{Wright}, E.~L., {Eisenhardt}, P.~R.~M., {Mainzer}, A.~K., {et~al.} 2010, AJ,
  140, 1868

\end{thebibliography}
% \end{thebibliography}
%

\end{document}